%% file: main.tex
\documentclass[10pt,conference]{IEEEtran}

\usepackage[T1]{fontenc}
\usepackage[utf8]{inputenc}
\usepackage{microtype}
\usepackage{graphicx}
\usepackage{dblfloatfix}
\usepackage{booktabs}
\usepackage{amsmath,amssymb}
\usepackage{algorithm}
\usepackage{algpseudocode}
\usepackage[table]{xcolor}
\usepackage{tabularx}
\usepackage{url}
\usepackage{cite}
\usepackage{balance}
\usepackage{hyperref}
\hypersetup{hidelinks}
\usepackage[capitalize]{cleveref}

  \usepackage[most]{tcolorbox}
\newtcolorbox{rqtab}[1]{
  enhanced, breakable, sharp corners,
  colback=white, colframe=black, boxrule=0.8pt,
  top=9pt, bottom=5pt, left=6pt, right=6pt, fontupper=\small,
  overlay={\node[fill=black, text=white, font=\bfseries\scriptsize,
    inner xsep=5pt, inner ysep=2pt, anchor=west]
    at ([xshift=10pt]frame.north west) {#1};}}
    
\renewcommand{\dh}{\textit{d2h}}
\newcommand{\snaptwo}{\textsc{snap2}}
\newcommand{\moot}{\textsc{moot}}
\newcommand{\synth}{\textsc{SynthCore}}
\newcommand{\bsllm}{\textsc{bs\_llm}}
\newcommand{\opro}{\textsc{opro}}
\newcommand{\snap}{\textsc{snap}} 

\newcommand{\ezr}{\textsc{ezr}} 
\newcommand{\rand}{\textsc{random}}
\newcommand{\rrp}{\textsc{rrp}}

\title{Better Together, in the Right Order: Classical-then-LLM Optimization for SE}

\author{%
  \IEEEauthorblockN{Srinath Srinivasan}
  \IEEEauthorblockA{NC State University\\ssrini27@ncsu.edu}
  \and
  \IEEEauthorblockN{Tim Menzies}
  \IEEEauthorblockA{NC State University\\timm@ieee.org}
}

\begin{document}
\maketitle

\begin{abstract} 
A growing body of work combines large language models (LLMs) with classical
optimizers for software engineering (SE) configuration tasks. Often, the
classical optimizer is in charge: it owns the search loop and calls the LLM only
to assist in subroutines (e.g. to warm-start the first generation, propose a
mutation, or stand in as a surrogate). We report that there is much value in the
reverse approach: seeding an LLM with the results from a cheap classical learner.

We call this method SNAP2. Applied to over 100 SE tasks, it is the single
best of all methods studied, reaching the top tier on \textbf{85\%} of tasks,
ahead of the same LLM run alone (\textbf{75\%}) and ahead of every method in
which the classical optimizer retains control. It is also less expensive:
relative to the LLM-alone method, it uses roughly \textbf{30\% fewer tokens} and
runs \textbf{1.4$\times$ faster}, since the classical setup performs the
inexpensive work, and the LLM is invoked only to finish.

We conclude that it is unwise to study classical learners or LLMs in isolation:
there is much value in combining the two, and in the order that combination is
applied.
\end{abstract}


\section{Introduction}
\label{sec:intro}

\input{data}

Misconfiguration is one of the most expensive recurring failures in software.
In \textsc{MySQL}, \textsc{Apache}, and \textsc{Hadoop}, over 40\% of failures
trace to configuration errors~\cite{zhou2011understanding}, as do most cloud
production outages~\cite{tang2023fail}. A single bad setting is not a marginal
cost: \textsc{Storm}'s worst configuration runs $480\times$ slower than its
best~\cite{jamshidi2016uncertainty}. The problem is growing. With LLMs and
``vibe coding'' it is easy to add ``just a few more options'', and modern
systems already ship with thousands that few developers
understand~\cite{xu2015hey}: in 15 years \textsc{Postgres} options grew 300\%
and \textsc{MySQL} 600\%~\cite{vanaken2017automatic}. Most go untouched, so
users fall back on sub-optimal defaults~\cite{xu2015hey}.

Twenty-five years of search-based software engineering have attacked this with
stochastic search, Bayesian optimization, and active
learning~\cite{nair2020finding,chen2021efficient,siegmund2015performance,hutter2011smac}.
Lately the field has tried to move some, or all, of that work onto large
language
models~\cite{yang2023opro,senthilkumar2024can,liu2024lmea,liu2024llambo,schwanke2026mohollm,zhang2025llmmoea,romera2024funsearch}, but
it cannot agree on whether the LLM helps at all. Reports range from limited
advantage over conventional heuristics~\cite{huang2024exploring}, to useful only
on small problems~\cite{senthilkumar2024can}, to useful only as a search
subroutine~\cite{schwanke2026mohollm}, to too slow to be
practical~\cite{ganguly2025low}.

\begin{figure}[t]
\centering
\begin{tikzpicture}[font=\scriptsize]
  \def\rx{3.0}\def\ry{2.3}
  \definecolor{snapslate}{HTML}{2D2A32}
  \fill[teal!18]   (-1.3,0) ellipse [x radius=\rx, y radius=\ry];
  \fill[violet!18] ( 1.3,0) ellipse [x radius=\rx, y radius=\ry];
  \draw[teal!65!black,thick]   (-1.3,0) ellipse [x radius=\rx, y radius=\ry];
  \draw[violet!65!black,thick] ( 1.3,0) ellipse [x radius=\rx, y radius=\ry];
  \node[teal!45!black,align=center,font=\footnotesize]   at (-1.3,2.75)
     {\textbf{Classical optimizer}\\\textbf{drives the search}};
  \node[violet!45!black,align=center,font=\footnotesize] at ( 1.3,2.75)
     {\textbf{LLM is the}\\\textbf{optimizer}};
  \node[align=center] at (-2.95,0)
     {\textit{opt3 $\cdot$ no LLM}\\ EZR $\cdot$ RAND $\cdot$ RRP\\[4pt]
      \textit{opt2 $\cdot$ assists}\\ bs\_llm\\ SynthCore\\ LLAMBO $\cdot$ LMEA\\ GPTuner};
  \node[align=center] at (2.95,0)
     {\textit{opt1 $\cdot$ LLM only}\\ OPRO\\ EvoLLM\\ SNAP};
  \node[fill=snapslate,rounded corners=2pt,inner sep=2.5pt,text=white] at (0,0.55) {\textbf{SNAP2}};
  \node[align=center] at (0,-0.35)
     {classical seeds\\ $\rightarrow$ LLM optimizes};
\end{tikzpicture}
\caption{Methods grouped by who controls the search. Left: a classical optimizer
drives the search (opt3, no LLM; or opt2, where the LLM only assists a
subroutine). Right: the LLM is the optimizer (opt1). \snaptwo{} alone occupies
the intersection: a cheap classical learner seeds the trajectory, then the LLM
performs the optimization. Every prior opt2 method runs the LLM \emph{first} and
lets a classical optimizer finish; \snaptwo{} reverses that order, and no prior
method occupies this cell.}
\label{fig:venn}
\end{figure}
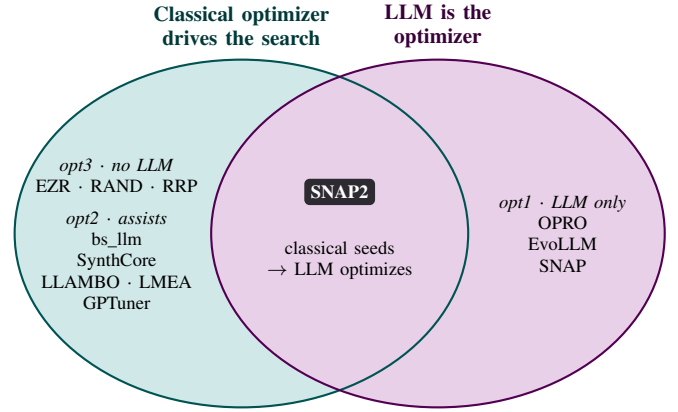

We argue this disagreement reflects an evidence base too thin to settle its own
disputes. Studies here often report five datasets or
fewer~\cite{senthilkumar2024can}, where per-dataset variance swamps the effect
sizes separating methods. Worse, papers reaching opposite verdicts on ``LLMs for
optimization'' may be measuring different regimes (\cref{fig:venn}). The phrase
covers at least three:
\begin{description}
  \item[{\bf opt1}] The LLM \emph{is} the only optimizer, re-prompting itself
        with the trajectory so far and no external search.
  \item[{\bf opt2}] A classical optimizer owns the search loop and the LLM
        assists in subroutines (e.g. warm-starting the first generation,
        proposing a mutation, or standing in as a
        surrogate)~\cite{senthilkumar2024can,liu2024lmea,liu2024llambo,schwanke2026mohollm,zhang2025llmmoea,romera2024funsearch}.
  \item[{\bf opt3}] A classical optimizer runs with no
        LLM~\cite{lindauer2022smac3,bergstra2011tpe,balandat2020botorch,zhang2007moead,bergstra2012random}.
\end{description}
Settling the disagreement therefore needs more evidence across regimes, not just
another algorithm. That is now possible: recently, the search based SE community collected over a
hundred real-world tasks into the \moot{} repository~\cite{menzies2025moot}
(\cref{mootdata}). Over 105 of those tasks we rank methods spanning all three
regimes (twenty repeats each) to ask:
\begin{description}
\item[{\bf RQ1:}] Does LLM involvement beat the classical state of the art?
\item[{\bf RQ2:}] Does combining a classical learner with an LLM beat either alone?
\item[{\bf RQ3:}] Does the order of combination matter?
\item[{\bf RQ4:}] What does each method cost, and who should run which?
\end{description}

Our results overturn the  pessimism reported above. Across opt2 work, the classical
optimizer is in charge and the LLM only assists, and crucially, the LLM acts
\emph{first}: it seeds, suggests, or prunes, then a non-LLM algorithm drives the
search to its answer. 

We find more value in the reverse approach: let a cheap classical
learner do the inexpensive early work, then hand its results to the LLM to
finish. This fills the middle cell of \cref{fig:venn}
(classical seeds, LLM
optimizes), a pairing no prior method occupies. We call it \snaptwo{}, and it is
the single best method we studied, reaching the top tier on \textbf{85\%} of
tasks: ahead of the same LLM run alone (\textbf{75\%}) and ahead of every method
in which the classical optimizer keeps control. It is also cheaper than the LLM
alone (roughly \textbf{30\% fewer tokens} and \textbf{1.4$\times$ faster}) since
the classical step does the inexpensive work and the LLM is invoked only to
finish.

We also find that the cheap classical learner itself, \ezr{}, while statistically
inferior to \snaptwo{}, still performs strongly and runs thousands of times
faster at zero token cost. \ezr{} alone may therefore be the right choice under
tight resource constraints, such as edge computing or restricted API budgets.

This paper makes four contributions:
\begin{enumerate}
\item {\em \snaptwo{}}, an optimizer that seeds an LLM with the output of a cheap
      classical learner, the reverse of the usual warm-start direction.
\item {\em A large-scale evaluation} across {\em opt1}, {\em opt2}, and {\em
      opt3} on 105 SE optimization tasks (\cref{sec:results}).
\item {\em Evidence that order matters}: combining a classical learner and an LLM
      beats either alone, and the cheap-classical-first ordering of \snaptwo{}
      beats the LLM-led ordering used in prior opt2 work.
\item {\em Repeatable, refutable science}: all code, prompts, and data are
      released under an open source license.
\end{enumerate}

\noindent
\cref{sec:background} motivates the problem for an SE audience;
\cref{sec:methods} describes our algorithms; \cref{sec:setup} covers the
\moot{} datasets, metrics, statistics, and labeling budget; \cref{sec:results}
answers the RQs; and \cref{sec:discussion,sec:threats} discuss results and
threats to validity.

\section{Background}
\label{sec:background}

\subsection{Is This an SE Problem?}
\label{sec:is-se}
As shown in \cref{mootdata}, the methods of this paper support a wide range
of common SE problems. All share one concern:
``find the best value for this slot''.
Formally, these are \emph{configuration tasks}, and SE has many:
\begin{itemize}
\item Effort estimation:   options for faster, better delivery;
\item Defect prediction: options that best locate defects;
\item Performance tuning: compile options for more speed, least energy;
\item Product lines: the best options per product;
\item Project health: running a project to attract committers;
\item Cloud allocation: the \texttt{.yaml} that best tunes the cloud;
\item Test-suite selection: most bugs, least runtime;
\item Cross-domain: old-project options that   help a new one;
\item Process modeling: steering a system-dynamics model to match the
      institutions and patterns that build the software.
\end{itemize}
Hand-tuned solutions to these problems are error-prone.
Hence, the failure costs of
\cref{sec:intro} are the direct consequence.

\subsection{Why These Problems Are Hard}
\label{sec:why-hard}

The difficulty is the size of the space a good answer hides in.
\textsc{x264}'s eleven parameters take roughly 1{,}536 hours of
compile-and-test~\cite{valov2017transferring}; \textsc{MySQL}'s 460 binary
options, treated as independent, yield $2^{460}$ combinations (the observable
universe holds only about $2^{80}$ stars)~\cite{xu2015hey}; \textsc{7z}'s
fourteen compression parameters generate about a million
configurations~\cite{chen2026promisetune}. Spaces this large must be pruned with
knowledge heuristics, and the three regimes of \cref{sec:intro} differ only in
where that knowledge comes from.

\subsubsection{opt1: only LLMs}
\label{sec:rel-opt1}
An LLM is a knowledge heuristic with no algorithmic search around it: it draws on
what it learned from terabytes of text, prompted to surface the part overlapping
the task. \opro{}~\cite{yang2023opro} is the canonical case: it feeds the model
the trajectory of tried points and asks for the next, with later
chain-of-thought variants~\cite{wei2022cot}.

\subsubsection{opt2: LLM plus classical}
\label{sec:rel-opt2}
Rather than discard decades of algorithmic work, opt2 mixes the two. The cleanest
case is the warm start: evolutionary optimizers need a generation~zero, and better
initial candidates yield better final results, so Senthilkumar et al.\ ask an LLM
to propose it~\cite{senthilkumar2024can,senthilkumar2026synthcore}.

\subsubsection{opt3: no LLM}
\label{sec:rel-opt3}
SMAC fits a random forest to the data seen so far and uses it to pick what to label
next, searching widely while uncertain, then exploiting the region the ensemble
agrees is best~\cite{hutter2011smac,lindauer2022smac3}. TPE does the same with two
Bayesian models~\cite{bergstra2011tpe}; multi-objective variants maximize expected
hypervolume (qEHVI~\cite{balandat2020botorch,daulton2021qehvi}) or decompose one
hard problem into many easy ones (MOEA/D~\cite{zhang2007moead}); DEHB adds
Hyperband's multi-fidelity budgeting~\cite{awad2021dehb}. \rrp{}, used here and
described below, is another no-LLM method.

We do not re-run these, and that omission is deliberate. At FSE'26, Ganguly et al.\
benchmarked SMAC3, TPE, and DEHB against \ezr{} on this same \moot{} corpus and found \ezr{} was
strongest~\cite{ganguly2025low}.

\begin{table*}[t]
\centering
\setlength{\tabcolsep}{5pt}
\caption{The seven methods on uniform axes. ``LLM role'' is how the model is
used. ``EZR substrate?'' is whether the method uses the \ezr{} active-learning
loop of \cref{sec:ezr} to choose rows to label. ``Warm start'' is how
generation~zero is seeded. $B$ is the label budget. ``Distinctive mechanism''
names each method's one new idea.
\snap{} and \snaptwo{} are new; the rest are adopted from prior work.}
\label{tab:methods}
\begin{tabular}{@{}llllcll@{}}
Method & Regime & LLM role & EZR substrate? & $B$ & Warm start & Distinctive mechanism \\
\midrule
\snaptwo & hybrid & finisher        & half & 20            & \ezr{}, $B{=}10$  & classical seeds, LLM finishes \\
\snap    & opt1   & sole optimizer  & no   & 20            & 4 random          & collision tracking \\
\bsllm   & opt2   & proposer        & yes  & 20            & LLM, one-shot     & whole-table one-shot prompt \\
\synth   & opt2   & proposer        & yes  & 20            & LLM, $M$ sessions & per-cluster session ensemble \\
\ezr     & opt3   & none            & yes  & 20            & random            & --- \\
\rand    & opt3   & none            & no   & 20            & random            & --- \\
\rrp     & opt3   & none            & no   & $\sim\!\log_2 n$ & rare-pair      & rare-pair projection \\
\bottomrule
\end{tabular}
\end{table*}

\section{Algorithms, the Details}
\label{sec:methods}

In SE optimization it is usually cheap to \emph{see} the space of options but
expensive to \emph{measure their effects}. For example:
\begin{itemize}
\item
A test suite of a billion cases is
easy to write and slow to run. 
\item Every parameter in a makefile is trivial to list,
but running them all to measure energy and runtime is often too slow to attempt.
\item
Analyst skill and tooling clearly lower the cost of good software, yet pinning a
number on that effect takes real work. 
\end{itemize}
So all our methods adopt one
meta-procedure, \emph{sequential model-based optimization} (SMO): given a table
split into $x$ (independent) and $y$ (dependent) columns, find rows with better
$y$ values while inspecting the $y$ values of at most $B$ rows. We call $B$ the
\emph{optimization budget}. It can be surprisingly small when SMO uses active
learning~\cite{settles2012active}. A passive learner builds its model from all
the data. An active learner instead reflects on the model built so far to pick
what to label next, so it can learn from very little data by chasing the most
informative example and skipping the redundant ones.

Every method here scores a candidate the same way. Each proposed configuration is
projected onto its nearest measured row, by feature distance $\rho$, so every
guess maps to a row with a known label. The methods differ only in which rows
they spend the budget on. We compare seven of them across the three regimes
(\cref{tab:methods}). 

The candidate space of {\em opt1,opt2,opt3} algorithms is far too large for one paper to
enumerate.
Accordingly,  for each regime we take the published state of the art and add our
own where the design space is empty. That coverage is deliberately uneven,
because the literature is uneven:
\begin{itemize}
\item
Opt1 is nearly bare: the field mostly assumed a
standalone LLM could not optimize~\cite{huang2024exploring}, leaving
\opro{}~\cite{yang2023opro} as the one landmark which we extend into \snap{}.
\item
Opt2
gives us two recent methods, \bsllm{}~\cite{senthilkumar2024can} and
\synth{}~\cite{senthilkumar2026synthcore}.
\item
opt3 gives us \ezr{}, the learner
Ganguly et al.\ recommend after a large FSE'26
evaluation~\cite{ganguly2025low}.
\end{itemize}
Cohen advises checking complex methods against
simple ones~\cite{cohen1995empirical}, so we add two no-LLM floors: \rand{} and
the 2018 search-based method \rrp{}~\cite{chen2018sway}.

Four of the seven algorithms are quickly described.
\begin{itemize}
\item \rand{} is the simplest. It randomly samples, labels, and sorts $B$ rows,
      then returns the best (by \cref{eq:delta}).
\item \rrp{} (random rare-pair) picks two far-apart rows 
      then projects every row onto the line between them (using the cosine
      rule\footnote{Given rows $A,B$ at distance $c$, a row $X$ at distances
      $a,b$ to $A,B$ projects onto the $A$--$B$ line at
      $x=\frac{a^2+c^2-b^2}{2c}$.}). It splits at the median and recurses on the
      half nearer the better pole. It labels two rows per split, so it needs
      $O(\log_2 n)$ labels.
\item \synth{}~\cite{senthilkumar2026synthcore} seeds by \emph{clustered}
      ensemble prompting. DBSCAN~\cite{ester1996dbscan} pre-clusters the labeled
      rows. Across $M$ independent sessions, each with a reset context and a
      fresh per-cluster sample, the model is asked to synthesize a row better
      than the best seen so far. There is no chain-of-thought, no debate, no
      cross-talk between sessions. The $M$ outputs are sorted and the best is
      returned.
\item \bsllm{} (``baseline LLM'') is an un-clustered
      version of \synth{}. One prompt over all the data guesses $B$ configurations. It is the
      naive warm-start proposer that the clustered \synth{} should beat.
\end{itemize}
The other three methods carry the paper's argument, so we detail them next.

\begin{algorithm}[!t]
\caption{\ezr{}: centroid-driven active learning, the no-LLM control. A
\emph{best}/\emph{rest} split is grown one label at a time; the centroids
$C_b,C_r$ pick the next row to label. The warm start is random (cf.\ the
LLM-seeded opt2 methods).}\label{alg:ezr}
\small
\begin{algorithmic}[1]
\Require rows $R$, warm-start size $W$, total budget $B$,
         feature distance $\rho$, objective score $\textit{d2h}$
\State shuffle $R$;\quad $S \gets R[1{:}W]$ \Comment{random warm start; label $S$}
\State sort $S$ \textbf{ascending} by $\textit{d2h}$ \Comment{lower is better}
\State $m \gets \lfloor\sqrt{W}\rfloor$
\State $\mathit{best} \gets S[1{:}m]$;\quad $\mathit{rest} \gets S[m{+}1{:}]$
\State $D \gets S$;\quad $U \gets R \setminus S$ \Comment{labeled / unevaluated}
\While{$|D| < B$ \textbf{and} $U \neq \emptyset$}
  \State $C_b \gets \textit{mids}(\mathit{best})$;\quad $C_r \gets \textit{mids}(\mathit{rest})$
  \State $r^\ast \gets \arg\min_{r \in U}\ \rho(r,C_b) - \rho(r,C_r)$ \Comment{nearer best}
  \State $U \gets U \setminus \{r^\ast\}$;\quad label $r^\ast$ \Comment{spend one of $B$}
  \State $D \gets D \cup \{r^\ast\}$;\quad $\mathit{best} \gets \mathit{best} \cup \{r^\ast\}$
  \If{$|\mathit{best}| > \lfloor\sqrt{|D|}\rfloor$} \Comment{rebalance}
    \State sort $\mathit{best}$ \textbf{ascending} by $\textit{d2h}$
    \State $\mathit{rest} \gets \mathit{rest} \cup \{\,\textsc{pop-worst}(\mathit{best})\,\}$
  \EndIf
\EndWhile
\State \Return $\arg\min_{r \in D}\ \textit{d2h}(r)$
\end{algorithmic}
\end{algorithm}

\subsection{\ezr{} (opt3): no LLM}
\label{sec:ezr}

\ezr{}~\cite{senthilkumar2024can,ganguly2025low} is our opt3 method and the
no-LLM control. It spends the same budget $B$ as the opt2 proposers but seeds
generation~zero at random instead of from an LLM (\cref{alg:ezr}). It labels $W$
random rows and partitions them, by \dh{} (\cref{metrics}), into a
$\sqrt{W}$-sized \emph{best} set and a \emph{rest} set with centroids $C_b,C_r$.
After that it labels rows one at a time. At each step it scores every unevaluated
row by
$\epsilon(\textit{row}) = \rho(\textit{row},C_b) - \rho(\textit{row},C_r)$,
where \emph{lower} is \emph{better} since it selects
for rows nearer \emph{best} and further from
\emph{rest}. The winning row is labeled and added to \emph{best}
(possibly   evicting the
worst member to \emph{rest} if \emph{best} grows past $\sqrt{|D|}$). This keeps
\emph{best} a small elite whose centroid sharpens as the budget is spent
(so each
label is chosen against a steadily better picture of the good region).

\ezr{} never queries a model and never invents a synthetic configuration. It only
\emph{selects} which tabulated row to reveal next, which makes it the natural
baseline method to compare against the LLM. It is also the exact
configuration Ganguly et al.~\cite{ganguly2025low} report beating the classical
state of the art (TPE~\cite{bergstra2011tpe}, SMAC3~\cite{lindauer2022smac3}, and
DEHB~\cite{awad2021dehb}) on these \moot{} tasks, so we adopt it unchanged.

\subsection{\snap{} (opt1): the LLM is the optimizer}
\label{sec:snap}

Algorithm~\ref{alg:snap} shows the \snap{} algorithm.
This is an opt1 algorithm; i.e.  the LLM \emph{is} the only optimizer so
there is no support by any classical optimizer.

\snap{}  is our extension to \opro{}~\cite{yang2023opro}. \opro{}'s main
loop shows the LLM a   trajectory of tried points and asks for the next.
\opro{} was validated only on continuous benchmarks (BBOB~\cite{hansen2009bbob},
TSP, vehicle routing), where Huang et al.\ found standalone LLMs no better than
conventional heuristics~\cite{huang2024exploring}. \moot{} data is largely
symbolic, so we adapt \opro{} to selection over a fixed table, as described here.

\begin{algorithm}[!t]
\caption{\snap{}/\snaptwo{}: \opro{}-style re-prompting with no external active
learner. A collision set \emph{coll} steers the prompt away from regions already
explored. \snap{} uses the random warm start ($v{=}1$); \snaptwo{} uses the
\ezr{} warm start ($v{=}2$).}\label{alg:snap}
\begin{algorithmic}[1]
\small
\Require rows $R$, warm-start size $W$, warm-start mode $v$, total budget $B$,
         candidates/call $K$, feature distance $\rho$, objective score $\textit{d2h}$
\State $D \gets \textsc{WarmStart}(R,\ W,\ v)$ \Comment{$|D|=W$}
\State $U \gets R \setminus D$ \Comment{unevaluated pool}
\State $T \gets \{(r,\ \textit{d2h}(r)) : r \in D\}$ \Comment{trajectory = optimizer state}
\State $\textit{coll} \gets \emptyset$
\While{$|D| < B$ \textbf{and} $U \neq \emptyset$}
  \State $k \gets \min(K,\ B - |D|)$
  \State sort $T$ \textbf{descending} by score \Comment{so best=T[0]}
  \State $C \gets \textsc{llm}(T,\ \textit{constraints},\ \textit{objectives},\ \textit{coll},\ k)$
  \State $\textit{coll} \gets \emptyset$
  \ForAll{$c \in C$}
    \State $r^\ast \gets \arg\min_{r \in U}\ \rho(c, r)$ \Comment{find near unevaluated }
    \State $r^\circ \gets \arg\min_{r \in D}\ \rho(c, r)$
    \If{$\rho(c, r^\circ) < \rho(c, r^\ast)$} \Comment{look elsewhere}
      \State $\textit{coll} \gets \textit{coll} \cup \{(r^\circ, \textit{d2h}(r^\circ))\}$
    \EndIf
    \State $U \gets U \setminus \{r^\ast\}$;\quad $D \gets D \cup \{r^\ast\}$
    \State $T \gets T \cup \{(r^\ast,\ \textit{d2h}(r^\ast))\}$
  \EndFor
\EndWhile
\State \Return $\arg\min_{(r,s)\in T} s$
\Statex
\Function{WarmStart}{$R,\ W,\ v$}
  \If{$v = 1$} \Comment{random ( \snap{} )}
    \State $D \subset R$ \quad \Comment{$|D|=W$}
  \ElsIf{$v = 2$} \Comment{ \ezr{} seed ( \snaptwo{} )}
    \State $D \gets \textsc{ezr}(R,\ B{=}10)$
  \EndIf
  \State \Return $D$
\EndFunction
\end{algorithmic}
\end{algorithm}

\begin{figure*}[t]
\scriptsize
\begin{verbatim}
### 1. SYSTEM MESSAGE
You are an optimizer. Your job is to propose feature configurations that minimize d2h
(distance to heaven, i.e.\ distance to the ideal objective values; lower is better). STRICT RULE: every value you propose MUST be
inside the declared NUM range or the declared SYM allowed-values set. Do NOT use values from your
prior knowledge of the domain - only the values listed in the constraints below are valid.

### 2. USER MESSAGE
## Hard constraints (each proposed value must satisfy these EXACTLY)
- Spout_wait (NUM): MUST be numeric, range [1, 10000]
- Spliters   (NUM): MUST be numeric, range [1, 6]
- Counters   (NUM): MUST be numeric, range [1, 18]

## 3. Objectives (NOT to be proposed; they are evaluated for you)
- Throughput+: maximize
- Latency-:    minimize

## 4. Trajectory   (sorted worst -> best; best is last, end-of-prompt)
| configuration                        | d2h    |
|--------------------------------------|--------|
| Spout_wait=1, Spliters=1, Counters=1 | 0.4681 |
| Spout_wait=1, Spliters=1, Counters=2 | 0.3806 |

## 5. Diversification (your prior proposals mapped onto these already-evaluated rows; propose distinct ones)
- Spout_wait=1, Spliters=1, Counters=4   (d2h=0.3582)

## 6. Retry  (your previous response was not valid: candidate 0 missing features: ['Counters'])
Prior response: {"candidates": [{"Spout_wait": 1, "Spliters": 2}]}
Re-read the hard constraints above. Every NUM must be in range; every SYM EXACTLY one of the
listed values.

## 7. Reminder
Verify each value against the hard-constraints block. NUM values outside [lo, hi] and SYM values
not in the allowed set will be rejected and waste a retry.

## 8. Instruction
Propose exactly 2 new feature configurations whose d2h is below 0.3806. Respond with ONLY a JSON
object of this exact shape (no prose, no markdown fences):
{ "candidates": [
    { "Spout_wait": <numeric in [1,10000]>, "Spliters": <numeric in [1,6]>,
      "Counters":   <numeric in [1,18]> },
    ...
] }
The "candidates" array MUST have exactly 2 entries.
\end{verbatim}
\caption{An example \snap{} prompt, on a 3-feature task. The eight blocks match
\snap{}'s prompt template: (1)~\textbf{System} sets the role and the strict
in-range rule; (2)~\textbf{Hard constraints} list each feature's legal range;
(3)~\textbf{Objectives} are described but never proposed (they are scored for the
model); (4)~\textbf{Trajectory} lists tried configurations sorted worst-to-best,
with the best placed last so it sits in the high-attention end-of-prompt
position; (5)~\textbf{Diversification} lists prior proposals that collided with
already-evaluated rows (the \textit{coll} set of \cref{alg:snap}), asking for
distinct ones; (6)~\textbf{Retry} feeds back the last malformed response so the
model can repair it; (7)~\textbf{Reminder} restates the validation rule; and
(8)~\textbf{Instruction} fixes the output to a JSON object of exact arity.}
\label{fig:snap-prompt}
\end{figure*}

Like \opro{}, \snap{} keeps a trajectory of tried configurations and their \dh{} scores. At
each step it asks for the next batch ($K{=}2$ per call) until the budget is spent
(\cref{alg:snap}). Its one addition to \opro{} is \emph{collision tracking}. When
a proposal lands nearer an already-evaluated row,  \snap{} nudges the suggestion
to another region.  \snap{} collects these ``nudges'' in a
set \textit{coll} and feeds it back as a ``diversification'' block
to the prompts of 
\cref{fig:snap-prompt}.
This pushes the model off exhausted neighborhoods with
no external search. The trajectory is seeded with $W{=}4$ random rows to match the
other methods. 
\snap{}'s contribution is subtractive: where the opt2 methods hand
their proposals to a downstream active learner, \snap{} deletes that stage and
lets the model choose every point.

\subsection{\snaptwo{} (the intersection): the classical learner seeds the LLM}
\label{sec:snap2}

\snaptwo{} was not designed up front. It fell out of the results of this paper. Once every
method had run (\cref{sec:results}), one number was hard to ignore: \ezr{} uses no
LLM and runs three orders of magnitude faster than \snap{}, yet it trailed
\snap{} by only a few top-tier points. 
To say that another way, a nearly free method was almost as good as our best, most expensive one. That gap suggested a trade: i.e. use a little of the cheap
method to save time on the more expensive one.

 As shown at the bottom of Algorithm~\ref{alg:snap},  \snaptwo{} replaces \snap{}'s random warm start with the output of a short
\ezr{} run. \ezr{} spends the first half of its labeling budget acquiring informative rows.
The labeled trajectory then becomes generation~zero for the \opro{}-style loop,
which spends the second half ($\textsc{WarmStart}$ mode $v{=}2$ in
\cref{alg:snap}). The total label budget stays at $B$ and nothing else changes.

\section{Experimental Setup}
\label{sec:setup}

\subsection{Datasets: \moot}

This study used the  MOOT data from \cref{mootdata}.
As shown in that table, these tabular data sets have
\begin{itemize}
\item 3-1000 independent $x$ attributes;
\item 1 to 8 dependent $y$ attributes to be minimized or maximized;
\item 93 to 100,000 rows.
\end{itemize}
To the best of our knowledge, MOOT is the largest collection of real
SE optimization tasks assembled to date.
As shown by the references (right-hand-side column,  \cref{mootdata}),
MOOT's data come
from    papers by numerous SE authors, presented in leading venues
(ICSE, FSE, TSE, IST, EMSE, TOSEM, ASE). MOOT's tasks span configuration,
performance tuning, product lines, project health, defect prediction,
testing, cost estimation, cross-domain generalization, and text mining.
Compared to MOOT, other resources (e.g., SPLOT) were narrow and are now offline. Toolkits
like Pygmo or Platypus offer only synthetic benchmarks. MOOT's datasets come
from published studies, real performance logs, cloud systems, and tuning
tasks where bad configurations cost time, money, and credibility.

Of MOOT's 127 tasks we keep the 105 on which every method completed its full run of
repeats. This common-denominator rule keeps the comparison fair: no method can inflate its
score by silently skipping datasets it cannot complete.
We use this same set of 105 throughout; the per-dataset counts in
\cref{tab:winrate} and elsewhere all share this denominator.

The 105 datasets cover software configuration (\textsc{x264},
\textsc{BDBC}, \textsc{PostgreSQL}, \textsc{LLVM}, \textsc{Storm}, and
others), software process modeling (POM3, a model of agile
requirements prioritization~\cite{port2008pom3}; XOMO, a Monte Carlo
simulator over the COCOMO suite of effort, risk, and defect
models~\cite{menzies2005xomo}), feature
models, project health, and hyperparameter optimization. 


\subsection{Metrics}\label{metrics}

We report three layers of performance metrics. The underlying per-run \emph{score} is
distance to heaven. We normalize each objective into $[0,1]$, then compute:
\begin{equation}\label{eq:d2h}
  \mathit{d2h}(y) = \sqrt{\sum_{i=1}^{k} y_i^2} \, / \, \sqrt{k},
\end{equation}
where $k$ is the number of objectives and $0$ is the \emph{utopia
point}: the ideal but usually unreachable point where every
objective hits its best value at once.
For {\em d2h}, {\em lower} values are   {\em better}~\cite{senthilkumar2024can}
since those solutions are closer to the ideal points\footnote{ For the reader familiar with the optimization literature, we note that there are many other ways to measure performance in multi-objective reasoning.
A recent IEEE TSE article by Chen et al. \cite{Li22} reviewed various multi-objective optimization performance measures like hypervolume, Spread, Generational Distance, and Inverted Generational Distance. Subsequent work by  Chen~\cite{chen2026promisetune}, as well as analogous
research by Ganguly~\cite{ganguly2025low} suggests that these measures are somewhat misguided. At least for SE applications, when learning what $x$ leads to best $y$ values, 
most data falls to a tiny corner of $x$ space and $y$ space. Hence, the issue is not how spread out is a search, nor what is the total volume of your search, but how
quickly can you prune the search space to find the small portion relevant to your current tasks. Hence, here, we  used {\em d2h} rather than  HV, spread, GD and IGD.}

To make \dh{} comparable across datasets we report a normalized
\emph{improvement} $\Delta$. For each dataset, let
$\textit{base}_{\min}$ and $\textit{base}_{\mathrm{med}}$ be the
minimum and median \dh{} over every row of the CSV. Then
\begin{equation}\label{eq:delta}
  \Delta = 1 - \frac{\textit{d2h} - \textit{base}_{\min}}
                    {\textit{base}_{\mathrm{med}} - \textit{base}_{\min}},
\end{equation}
reported as a percent.
For $\Delta$, {\em larger} values are   {\em better}~\cite{senthilkumar2024can}
since those solutions are closer to the   best possible results:
$100\%$ means the empirical optimum
was recovered, $ 0\%$ means the no-tuning median, and
$\Delta > 100\%$ is possible when a method's held-out validation score
beats the best row in the table.

We will often report our results via the  \emph{top-tier rate}: the fraction of
datasets on which a method falls in the top tier, where ``best'' is defined in \cref{stats}.
Because top-tier membership is itself a significance decision (\cref{stats}), this rate
aggregates 105 per-dataset significance tests rather than raw scores. $\Delta$ gives the
magnitude of the improvement; the top-tier rate gives how often a method is among the best.

For cross-comparability with the evolutionary computation (EC)
literature, where multi-objective results are typically reported
through hypervolume or Chebyshev distance, we also computed both
metrics; the ranking is consistent, so we report \dh{}-based results
throughout.

\subsection{Cost Metrics}

For each (method, dataset, seed) triple, we measure five cost metrics: labels used, LLM tokens (input plus output), wall-clock seconds, number of LLM calls, and dollar cost in our setup.

\subsection{Statistical Analysis}\label{stats}

We ran twenty seeds per (method, dataset) pair. On each dataset, methods are
sorted by median score; the leader (best median) plus all subsequent methods
indistinguishable from it form the \emph{top tier} (stopping at the first method
that differs). Membership in this top tier is therefore a per-dataset
significance decision, not a raw score comparison: a dataset may place several
methods in its top tier, all credited equally. The \emph{top-tier rate} we
report (\cref{tab:winrate}) thus aggregates 105 such per-dataset significance
tests. To avoid significant-but-meaningless
splits~\cite{kampenes2007}, two result sets $X,Y$ (sizes $n,m$) are judged
indistinguishable if both of the following hold:
\begin{itemize}
\item Cliff's
      delta~\cite{cliff1993},
      $|\delta| = |\#(x{>}y) - \#(x{<}y)|\,/\,(nm)$,
      is below 0.195 (small~\cite{hess2004}); and
\item the
      Kolmogorov-Smirnov statistic~\cite{massey1951} (max distance between
      the empirical cumulative distributions) is below
      $1.36\sqrt{(n+m)/(nm)}$ (95\% confidence).
\end{itemize}
These two standard, distribution-free tests (a small effect size and no
significant difference) together decide whether two result sets differ.

\subsection{LLM and Hyperparameters}

All our LLM methods use \texttt{openai/gpt-oss-120b}\footnote{Why this model? We picked it for practical reasons, not novelty. It is inexpensive to run, popular and well served across providers, and widely used in the community. It is also open-weights, which matters most for replicability: other groups can mirror the exact model and reproduce these results without depending on a single vendor's API staying available.}
served through
OpenRouter, with temperature $0.7$, top-$p$ $1.0$, a 32K context
window, and a 4000-token output cap (\snap{} raises this to 8000 on
datasets with $\geq 77$ features). It is a reasoning model: roughly
half of every completion token is an unshown reasoning token, billed
at the output rate and included in the costs we report. Main runs used
a deterministic disk cache; the cost-timing run disables it for real
latency. Seeds were derived from a fixed base ($12345$) per repeat and
dataset and control Python-side randomness only.

\definecolor{rqhl}{HTML}{FFF2CC}  
\definecolor{hlq}{HTML}{D9EAD3}   
\definecolor{hlc}{HTML}{D9E2F3}   
\definecolor{hld}{HTML}{F4D9E6}   
\newcommand{\hl}[2]{{\setlength{\fboxsep}{1.5pt}\colorbox{#1}{#2}}}
\newcommand{\hlnum}[1]{\hl{rqhl}{#1}}


A key hyperparameter here is the labeling budget. Each LLM labeling incurs a
compute cost, so for fairness we fix the same labeling budget across all tools,
LLM and otherwise. (\snaptwo{} spends the first 10 of its 20 labels inside
\ezr{}, then the rest in the \snap{} loop, so its total matches the others.)
That budget is bounded by financial resources. University
cloud clusters often run older kernels that cannot host the latest models, so these
studies were run on local machines with compute purchased at commercial rates.
Consider a typical research group of five graduate students, each delivering three
reports per year. Each report consumes up to six full LLM runs:
\begin{itemize}
\item run one, to survey the options;
\item run two, to debug the scripts and prompts for the substantive analysis;
\item run three, to execute that substantive analysis;
\item runs four, five, and six, to repeat experiments that prior runs invalidated.
\end{itemize}

\begin{table}[!t]
\footnotesize
\caption{Ganguly et al.~\cite{ganguly2025low} argue that with sufficient background knowledge, (e.g. from an LLM),
it is possible to find solutions statistically
indistinguishable from the best solution with as few as $B\le 20$ samples in practice. }\label{proof}
\begin{tabular}{|p{.95\columnwidth}|}\hline
Hamlet~\cite{hamlet1987probable} comments on the number of samples $B$ required to be $C$ confident
of finding a target with probability  $p$ is  \[C=1-(1-p)^B \rightarrow B=\frac{\log(1-C)}{\log(1-p)}\]
With some background knowledge (e.g. an LLM) we can compare two
events and heuristically guess if one is better than another. This means that 
assuming
solutions are spread randomly over one dimension (e.g. the range of values seen in \cref{eq:delta}), then a binary chop can isolate the best
results in $\log_2(B)$ comparisons.
Cohen comments that events are indistinguishably different if they differ
by less than $.2\sigma$~\cite{sawilowsky2009new}. For a Gaussian of range
\mbox{$-3\sigma$ to $+3\sigma$}, this
0.2 is a target of size $p=.2/6\approx 0.033$. Combining all this, an
unguided search needs $B\approx 137$ samples to be 99\% certain of hitting that
target, but a binary chop with background knowledge isolates it in $\log_2(B)<8$
comparisons:
\begin{equation}\label{eq:co} 
B(C=0.99, p=0.033)=\left(\log_2\left(\frac{\log(1-C)}{\log(1-p)}\right)\right)< 8
\end{equation}
For many reasons, this analysis is overly optimistic. It rests on numerous
assumptions: a uniform spread of good solutions, independence between draws, and the
Gaussian normality premise behind Cohen's rule, none of which hold exactly for our
data. Empirically, too, \cref{eq:co} understates the cost: the optimizers need $B=20$
samples, not 8, to find solutions within 80\% of the optimum (median), and even then do
not reach near-perfect results. Clearly, real-world data is far quirkier than the
idealized settings assumed by Cohen, Hamlet, and colleagues. 

That said, given
 \cref{eq:co},
it is  plausible that   $B=20$ 
labels is a reasonable   choice for this study.\\\hline
\end{tabular}
\end{table}

We view LLMs as assistants, not replacements for human researchers.
Hence, our  annual LLM budget must stay
below the cost of a graduate research assistant (\$2,000 per month, plus 50\%
overhead). Across five students and three reports each, that caps every study
like this one at
\[\left(12 \times 1.5 \times \$2{,}000\right) / 6 / 5 / 3 = \$400.\]
This \$400 is spread across all the optimizers, and all examples must be
processed. In those terms, we need a labeling budget of $B=20$ per optimizer per
run, rising slightly for occasional ablation runs ($B \in \{5, 10, 20,
50\}$). 

So is it useful running optimizers with
labeling budgets as low as   $B=20$?
Just to state the obvious: a budget of $B=20$ labels would be inadequate for replacing an LLM. But the LLM methods of this paper are more
about alternatives to  
few-shot prompting rather than LLM creation.

Also,
 the methods of this paper are  active learners; i.e. they reflect on the results seen so built
so far in order to select the next example to label.  By focusing on the most informative examples, our learners can avoid redundant and spurious  items and build
models with very few labeled examples.  
Also, \cref{proof}
offers a  mathematical argument from Ganguly et al.~\cite{ganguly2025low}
that, with sufficient  knowledge  (e.g. from an LLM),
it is possible to find good solutions  with $B\le 20$.

\begin{table}[t]
\centering 
\caption{Top-tier counts across the 105 MOOT data sets. A method scores a
\emph{top} on a data set when it is statistically indistinguishable from the best
method there, a per-dataset significance test (see \cref{stats}). The top-tier
rate aggregates these 105 significance decisions.}
\label{tab:winrate}
\begin{tabular}{@{}r l r  r@{}}
\toprule
Method & Regime & Tops &  Top-tier rate (tops/105) \\
\midrule
\snaptwo & hybrid & 89 & 85\% \\
\snap     & opt1 & 79 &   75\% \\
\bsllm   & opt2& 78 &   74\% \\
\ezr    & opt3& 75 &   71\% \\
\synth  & opt2 & 73 &   70\% \\
\rand     & opt3&51 &   49\% \\
\rrp      & opt3 & 34 &  32\% \\
\bottomrule
\end{tabular}
\end{table}


\section{Results}
\label{sec:results}

Under the conditions above, we ran the seven optimizers on 105 data sets, twenty
repeats each. Results were scored with \cref{eq:delta} and compared using the
statistics of \cref{stats}. We organize the findings as four questions. RQ1 asks
whether any LLM involvement beats the classical state of the art. RQ2 and RQ3
correspond to the two claims in our title: that combining a classical learner with
an LLM beats either alone, and that the order of combination matters. RQ4 reports
what each method costs and who should run which.

\noindent{\bf RQ1}: {\em Does LLM involvement beat the classical state of the art?}

\cref{tab:winrate} sorts our results. Recall that, for each data set,
our statistical method sorts each method by its mean results. Then, it asks which methods are ``top-tier''; i.e. which are statistically indistinguishable from the best method in that set. The numbers in \cref{tab:winrate} show how often, across all 105 data sets, a method is top-tiered.

The clear losers are  the ``no knowledge'' methods.
\rand{} (49\%) and \rrp{} (32\%) use neither an LLM nor an active learner, and
every method with a knowledge source beats them. \ezr{} (71\%) is the strongest
no-LLM method and our classical anchor
(and recall that Ganguly et al.\ showed \ezr{} also beats many classical methods such as TPE,
SMAC3, and DEHB on these same tasks~\cite{ganguly2025low}). The LLM-bearing methods \snap{} (75\%) and \bsllm{} (74\%) sit just above \ezr{}, and \synth{}
(70\%) very nearly ties with it.

\begin{figure*}[!b]
\centering
\includegraphics[width=.9\textwidth]{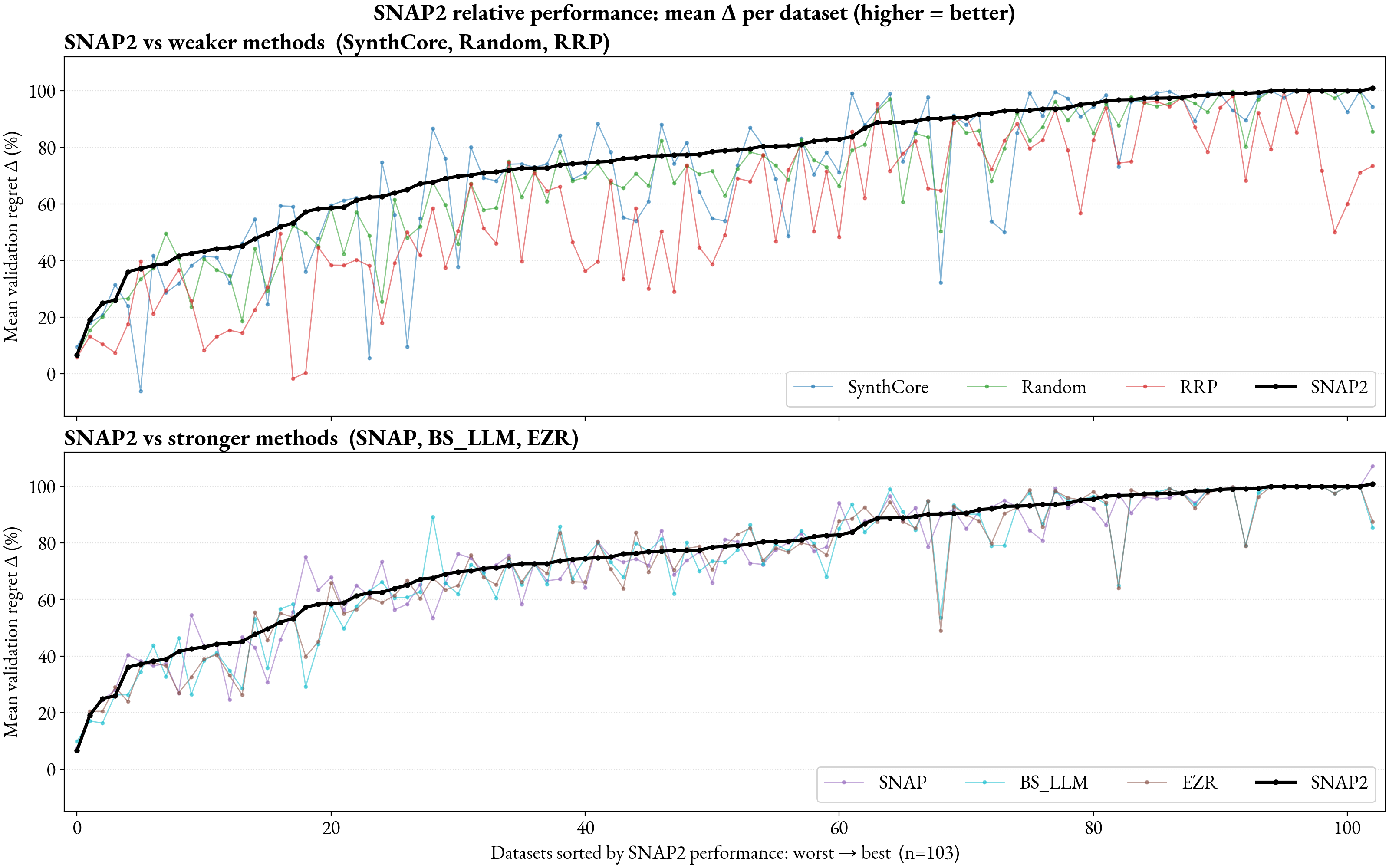}
\caption{Per-dataset mean $\Delta$ (higher is better, see \cref{eq:delta}) across the
103 of 105 datasets with a defined $\Delta$ baseline (2 have none), sorted left to right
by \snaptwo{} performance. Black line is
\snaptwo{}. \textbf{Top:} \snaptwo{} against the weaker methods (\synth, \rand, \rrp);
\snaptwo{} sits above them on most datasets. \textbf{Bottom:} \snaptwo{} against the strongest competitors
(\snap, \bsllm, \ezr).}
\label{fig:snap_relative}
\end{figure*}
 
So the answer to RQ1 is  ``yes'', with one caveat: the leading group is statistically
clustered. These four methods span only five points (70 to 75\%), inside the band
our per-dataset tests cannot separate (\cref{stats}). No single LLM-assisted method
pulls clear of \ezr{}. 

One method is left out of the above analysis: \snaptwo{}, the subject of RQ2 and RQ3.

\begin{rqtab}{RQ1}\textbf{Does LLM involvement beat the classical SOTA? Yes, but the results
cluster.} The knowledge-free floors (\rand{} 49\%, \rrp{} 32\%) are dominated.
Every LLM-bearing method matches or beats \ezr{} (71\%), which already beats TPE,
SMAC3, and DEHB~\cite{ganguly2025low}. But \snap{}, \bsllm{}, \synth{}, and \ezr{}
fall within a five-point band our tests cannot really separate. The open question is what
breaks that cluster (see RQ2).\end{rqtab}

\noindent{\bf RQ2}: {\em Does combining a classical learner with an LLM beat either
alone?}

\snaptwo{} combines the two: \ezr{} spends the first ten labels, then \snap{}
spends the rest. It reaches the top tier on 85\% of tasks (89 of 105), clear of
both ingredients alone (\snap{} 75\%, \ezr{} 71\%). It is the only method to break
the 70-to-75\% cluster of RQ1, by a margin our tests can separate.

\begin{rqtab}{RQ2}\textbf{Does combining beat either alone? Yes.} \snaptwo{}
reaches 85\%, above \snap{} alone (75\%) and \ezr{} alone (71\%). It is the only
method to escape the RQ1 cluster.\end{rqtab}

\newpage \noindent{\bf RQ3}: {\em Does the order of combination matter?}

The opt2 family (\bsllm{}, \synth{})   combines classical and LLM, but in the
opposite order to \snaptwo{}
(the LLM seeds, and a classical learner then drives the search).
\snaptwo{} reverses this, seeding with \ezr{} and finishing with \snap{}. The
reversal is decisive. \snaptwo{} (85\%) tops every LLM-first method by eleven
points or more (\bsllm{} 74\%, \synth{} 70\%), a gap outside the cluster band of
RQ1.  

\begin{rqtab}{RQ3}\textbf{Does order matter? Yes.} The LLM-first opt2 methods reach
74\% (\bsllm{}) and 70\% (\synth{}). Classical-first \snaptwo{} reaches 85\%.  \end{rqtab}

\noindent{\bf RQ4}: {\em What does each method cost, and who should run which?}

\cref{tab:cost} reports cost per repeat. The no-LLM methods make no API calls, so
they cost almost nothing and run very fast. \snaptwo{} dominates pure \snap{} on
every measure: a higher top-tier rate (85\% vs 75\%), about 30\% fewer tokens
(19{,}000 vs 26{,}000), $1.4\times$ faster (110.9\,s vs 157.7\,s), and roughly half
the dollar cost. Pure \snap{} is therefore never the right choice.

The live trade-off is \snaptwo{} against \ezr{}. The top-tier rates differ (85\% vs
71\%), and our tests call that difference significant. But the two are close in
\emph{absolute} mean performance (mean $\Delta$ 75.8\% vs 73.0\%, a 2.8-point gap;
\cref{fig:snap_relative}), while \ezr{} costs
nothing and runs three orders of magnitude faster (0.037\,s vs 110.9\,s). So the
small statistical gap can be worth conceding. 
We recommend running \snaptwo{} for mission- or
safety-critical work where quality is worth the dollars and latency. 
On the other hand, \ezr{} might be useful
when token budgets are capped, when response time is critical, or compute is scarce (e.g. for edge computing).

\begin{rqtab}{RQ4}\textbf{Who should run which?} \snaptwo{} dominates pure \snap{}
on quality (85\% vs 75\%) and cost, so pure \snap{} is never the right choice. Against \ezr{}, the
  top-tier gap is significant in relative terms (85\% vs 71\%)
  but small in absolute mean performance
(mean $\Delta$ 75.8\% vs 73.0\%; \cref{fig:snap_relative}). Also,  \ezr{} is virtually  free and far faster. Run \snaptwo{} when
quality is worth the cost. Run \ezr{} when budget, latency, or compute is the
binding constraint.\end{rqtab}

\begin{table}[t]
\centering
\caption{Operational cost per repeat, from  
\texttt{gpt-oss-120b} with cache disabled. }
\label{tab:cost}
\begin{tabular}{@{}l l r r r r@{}}
\toprule
Method & Type & LLM calls & Tokens & Cost (\$) & Wall (s) \\
\midrule
\rand     & opt3 & 0    & 0      & 0       &  0.0001 \\
\ezr  & opt3 & 0    & 0      &  0 & 0.037  \\
\rrp      & opt3 & 0    & 0      & 0       & 0.227  \\
\bsllm    & opt2 & 1  & 1{,}800  & 0.0005 & 29.5  \\
\synth    & opt2 & 1  & 2{,}000  & 0.0005 & 28.2  \\
\snaptwo & hybrid & 5 & 19,000 & 0.0039 & 110.9
\\
\snap     & opt1 & 8  & 26,000 &  0.0076 & 157.7 \\
\bottomrule
\end{tabular}
\end{table}

\section{Discussion}
\label{sec:discussion}

\subsection{Threats to Validity}
\label{sec:threats}

\paragraph{Dataset Selection}:  are the datasets a representative sample of all SE?

Every empirical study faces this threat to validity. The best we can do is make our case study library as large as possible, which is why we used MOOT.
The 105 datasets used here are the subset of MOOT's 127. The excluded datasets are method timeouts, not low-quality data. A
method that fails more often is therefore filtered out silently. The datasets are
also all tabular. Whether the conclusions transfer to non-tabular SE optimization
(test generation, patch synthesis, code optimization) is an open problem.

\paragraph{Construct Validity}: do our evaluations measure what we think they measure?

$\Delta$ is normalized against the empirical minimum and median \dh{} over the
dataset rows, not against an external ground truth. Here ``optimal'' means
\emph{reference optimal}, a term from the empirical-algorithms
literature~\cite{mcgeoch2012guide,cohen1995empirical}: the best solution observed so
far. The reference optimal may not be the true optimal. But in many engineering
cases the true optimal is unknown. Because it is only the best \emph{observed} row,
a method's held-out score can exceed it. This is why $\Delta$ can pass $100\%$, as
\snaptwo{} does on \textsc{pom3a} ($100.9\%$). We also computed hypervolume and
Chebyshev distance, and the ranking is consistent.

\paragraph{Shared Machinery}: does \snaptwo{} win only because it reuses a
baseline?

\snaptwo{} uses \ezr{} as its first stage, then is compared against
\ezr{} in the same ranking. The two are not independent, so \snaptwo{}'s advantage
could reflect that shared substrate rather than the classical-then-LLM order. Three
points bound this. The split was fixed in advance, not searched for the best result.
And \snaptwo{} also tops the LLM-first opt2 methods, which share \snap{}'s machinery
but reverse the order. So the order, not the shared parts, is what moves the
ranking. A third point is decisive. \snaptwo{} gives its \ezr{} stage only $B{=}10$
labels, half the $B{=}20$ that the standalone \ezr{} baseline gets (\cref{tab:methods}).
Its \ezr{} stage is therefore a weaker \ezr{} than the baseline, yet \snaptwo{} still
wins. The gain cannot come from the \ezr{} stage alone.

\paragraph{Sample Validity}: did we test enough of the design space?

We explore
some algorithmic options, but not all. Some omitted options were already assessed at
FSE'26, where \ezr{} beat SMAC3, TPE, and the qEHVI/MOEA/D
family~\cite{ganguly2025low}. The other omissions are the opt2 sub-roles: variation
operator~\cite{liu2024lmea,brahmachary2025leo}, surrogate~\cite{liu2024llambo},
caged sampler~\cite{schwanke2026hollm,schwanke2026mohollm}, operator
selection~\cite{zhang2025llmmoea}, and algorithm
designer~\cite{romera2024funsearch,liu2024eoh,ye2024reevo,gao2024sbllm}. These are
excluded by design. We hold the host optimizer fixed (the shared SMO loop at $B=20$)
and vary only the degree of LLM involvement. Warm-start seeding is the one sub-role
that swaps cleanly onto that loop. Each other sub-role presupposes a different host:
an evolutionary loop, a Bayesian surrogate loop, a partition scheduler, or a
program-synthesis loop. Importing them would confound the regime effect with a
host-optimizer effect. Several also abandon the fixed-$B$ label axis, which makes
them incommensurable in our ranking and cost tables. Testing which scaffold earns
its keep is left to future work.

\paragraph{External Validity}: do the results generalize beyond our setting?

\snap{} and \snaptwo{} propose configurations, then project each onto the nearest
\emph{measured} row. They select from a known table rather than generating unseen
configurations. This bounds external validity. Our results speak to choosing among
configurations whose objective values are already recorded, not to proposing
configurations never evaluated. Extending this approach to generative settings,
where the optimum may lie off-table, is future work.

\subsection{Open Issues and Future Work}
\label{sec:future}

\paragraph{A complexity measure for SE optimization.} The ceiling argument above
turns the usual question around. Instead of ``which optimizer is best,'' ask
``which problems are hard.'' Most \moot{} tasks yield to $B=20$ labels
(\cref{sec:results}); a few resist. A predictor of that resistance, computed from
cheap surface features of a dataset before any optimizer runs, would let a
practitioner budget labels per task. It would also explain why so many methods
converge. The landscape-sparsity intuition behind our $B=20$ argument
(\cref{proof}) is the natural starting point.

\paragraph{A conditional cascade.} \snaptwo{} runs \ezr{} then \snap{} on every
task. But \ezr{} alone already reaches most of \snaptwo{}'s quality at near-zero
cost (\cref{tab:winrate,tab:cost}). The open question is when the \snap{} stage is
worth running at all. A conditional cascade would run \ezr{} first, then escalate
to \snap{} only on tasks whose \ezr{} result looks uncertain. Paired with the
complexity measure above, this could capture \snaptwo{}'s quality while paying the
LLM cost only where it earns its keep.

\paragraph{Beyond tabular selection.} Our methods snap each proposed candidate to the nearest row in the \moot{} table, a
configuration whose objective values we have already measured. This snapping   means the LLM only has to point toward a good region, not name an
exact valid configuration, and it can never return an unmeasured or invalid answer.
But it only works because \moot{} gives us a finite table of scored configurations
to snap onto. Other SE optimization tasks have no such table. Test generation, patch
synthesis, and code optimization all search an open space, where no candidate is
pre-scored and there is nothing to snap to. Whether our results hold once that safety
net is removed, and what should replace it, is the main question this paper leaves
open.

\paragraph{Scaling with the model.} Our results come from one open-weights model,
one seed family, and $B=20$. As models improve, does \snaptwo{}'s margin over
\ezr{} widen, and does \ezr{} remain the stable free anchor? Half of the LLM tokens
are unshown reasoning tokens (\cref{sec:setup}). A study of cheaper, non-reasoning
prompting would show how much of the advantage is reasoning and how much is the
projection trick.

\section{Conclusion}
\label{sec:conclusion}

The literature was too pessimistic about LLMs as optimizers for SE. On 105 SE tasks,
the best method is neither a classical learner nor an LLM alone, but the two combined.
\snaptwo{} seeds a short \ezr{} run, then hands its trajectory to an LLM to finish. It
reaches the top tier on 85\% of tasks, above the same LLM run alone (75\%) and above
\ezr{} alone (71\%).  

The order of that combination is what wins. Prior work runs the LLM first and lets a
classical optimizer finish; those methods top out at 74\%. \snaptwo{} reverses the
order, seeds with the cheap classical learner, and lets the LLM close. Same
ingredients, opposite order, and an eleven-point gain (to 85\%). 

The result also costs less. \snaptwo{} dominates the LLM-alone method on every axis:
higher quality, about 30\% fewer tokens, and $1.4\times$ faster, since the classical
step does the cheap early work.

To conclude, we offer the following
guidance to  practitioners.  When optimizing
some SE process:
\begin{itemize}
\item
First try a classical
method (i.e. no LLM). Why? Well:
\begin{itemize}
\item
Some of those classical methods,
like \ezr{}, are virtually free and perform close to the best known results (yet run three orders of magnitude faster).
\item
By first running a very fast classical optimizer, practitioners can get quick feedback. 
\item
Such classical methods offer a baseline result against which the value of supposedly more valuable systems can be compared.
\item
Further, you will need the classical optimizer for the next step.
\end{itemize}
\item
Second, try an LLM-based optimizer  when developing (a) mission-critical
applications or (b) safety-critical applications, or (c) when  execution cost is not a concern. But first, bootstrap the LLM's search with the results of
a classical method. For a worked
example of how to do that, see
the 
\snaptwo{} implementation in this paper.
\end{itemize}

\balance
\bibliographystyle{IEEEtran}
\bibliography{main}

\end{document}

%% file: data.tex
\begin{table*}[!t] 
{\fontsize{7}{8}\selectfont 
\centering
\setlength{\tabcolsep}{2.5pt} 
\renewcommand{\arraystretch}{1.1} 

\caption{Experimental Datasets: the 127 MOOT benchmarks. Optimization objectives ($y$), features ($x$), and identifiers used in this study. Experiments run on the 105 on which all seven methods clear the coverage gate (Section~\ref{sec:setup}).}\label{mootdata}

\begin{tabularx}{.8\textwidth}{@{} r |  l l l X c | c | c | c @{}} 

\textbf{\#} & \textbf{Category} & \textbf{Focus} & \textbf{File Name(s)} & \textbf{Here, optimizers struggle with issues like...} & \textbf{\#y} & \textbf{\#x} & \textbf{Rows} & \textbf{Refs} \\ 

\multicolumn{9}{l}{\cellcolor{red!15}\textbf{Software Systems \& Configuration \& Tuning}} \\
25 & Soft. Config & \textbf{PLE} & SS-[A-X], billing10k & Minimize footprint/memory vs. maximizing throughput & 2-3 & 3--88 & 197–86k & \cite{Amiraliminimaldata} \\
\rowcolor{gray!10}
12 & PromiseTune & \textbf{Perf.} & 7z, LLVM, BDBC & Minimize execution time and energy consumption & 1 & 6--35 & 864–166k & \cite{chen2026promisetune} \\ 
1 & Cloud compute& \textbf{Tuning} & Apache\_AllMeas & Balance server response vs. CPU/RAM load patterns & 1 & 9 & 192 & \cite{senthilkumar2024can} \\ \rowcolor{gray!10}
1 & Cloud  compute& \textbf{Tuning} & SQL\_AllMeas & Minimize latency/IO vs. maximizing throughput & 1 & 39 & 4,654 & \cite{chen2025accuracy} \\ 
1 & Cloud compute & \textbf{Tuning} & X264\_AllMeas & Optimize encoding parameters for PSNR/SSIM quality & 1 & 16 & 1,153 & \cite{Amiraliminimaldata} \\ \rowcolor{gray!10}
1 & Cloud compute & \textbf{Tuning} & HSMGP\_num & Minimize solver runtime across grid configurations & 1 & 14 & 3,457 & \cite{chen2025accuracy} \\
7 & Cloud compute & \textbf{Tuning} & rs-*, sol-*, wc-* & Misc.\ configuration tuning trade-offs & 1 & 3-6 & 196–3,840 & \cite{senthilkumar2024can} \\ \rowcolor{gray!10}
2 & Testing & \textbf{Testing} & test120, 600 & Maximize coverage while minimizing execution time & 1 & 9 & 5,161 & -- \\ 

\multicolumn{9}{l}{\cellcolor{red!15}\textbf{Project Management \& Process Modeling}} \\
35 & Proj. Health & \textbf{Health} & Health-Closed & Optimize PR rates and minimize developer churn & 2-3 & 5 & 10,001 & \cite{lustosa2024learning} \\ \rowcolor{gray!10}
3 & Scrum & \textbf{Agile} & Scrum[1k-100k] & Maximize velocity within sprint constraints & 3 & 124 & 1k–100k & \cite{lustossa2024isneak} \\ 
8 & Feature Mod. & \textbf{Config} & FFM, FM-* & Optimize Clause/Constraint ratio in large spaces & 3 & 128--1k & 10,001 & \cite{Amiraliminimaldata} \\ \rowcolor{gray!10}
1 & nasa93dem & \textbf{Cost} & nasa93dem & Minimize effort (person-months) vs. quality & 3 & 26 & 93 & \cite{lustosa2024learning} \\ 
1 & COC1000 & \textbf{Risk} & coc1000 & Minimize risks/schedule slip vs. analyst expertise & 5 & 20 & 1,001 & \cite{chen2018beyond} \\ \rowcolor{gray!10}
4 & POM3 & \textbf{Process} & pom3[a-d] & Balance project idle rates vs. completion costs & 3 & 9 & 501–20k & \cite{lustossa2024isneak} \\ 
4 & XOMO & \textbf{Defects} & xomo\_[flt,grd] & Minimize defect density vs. total project effort & 4 & 27 & 10,001 & \cite{chen2018beyond} \\ 

\multicolumn{9}{l}{\cellcolor{red!15}\textbf{Interdisciplinary \& Behavioral Benchmarks}} \\ \rowcolor{gray!10}
4 & Behavioral & \textbf{Behavior} & all\_players, etc & Maximize user retention and predict cohort churn & 1-3 & 26--55 & 82–17k & \cite{nyagami_fc25_kaggle_2025} \\ 
4 & Financial & \textbf{Finance} & BankChurners & Minimize credit risk vs. optimal pricing models & 2-5 & 19--77 & 1k–20k & \cite{blastchar_telco_customer_churn_2025} \\ \rowcolor{gray!10}
3 & Health & \textbf{Health} & COVID19, Life & Maximize accuracy/life vs. readmission likelihood & 1-3 & 20--64 & 2k–25k & \cite{kumarajarshi_life_expectancy_who_2025} \\ 
2 & RL & \textbf{Control} & A2C\_[Acr,Crt] & Maximize reward signals vs. steps to convergence & 3-4 & 9--11 & 224–318 & -- \\ \rowcolor{gray!10}
5 & Sales & \textbf{Market} & Marketing, socks & Maximize ROI vs. minimizing forecasting error & 1-8 & 20--31 & 247–2k & \cite{jackdaoud_marketing_data_2022} \\ 
3 & Misc. & \textbf{General} & auto93, Car & Optimize multivariate trade-offs (e.g. MPG vs. HP) & 2-5 & 5--38 & 205–1k & \cite{Amiraliminimaldata} \\ \midrule

\textbf{127} & \multicolumn{8}{l}{\textbf{Total}}
\end{tabularx}
\par
}
\end{table*}